\documentstyle[12pt,epsfig]{article}
\begin{document}
\title{\vspace{-15mm}
       {\normalsize \hfill
       \begin{tabbing}
       \`\begin{tabular}{l}
         hep--th/9312167 \\
         SLAC--PUB--6411  \\
         December 1993 \\
         T 
         \end{tabular}       
       \end{tabbing} }
       \vspace{8mm}
\setcounter{footnote}{1}
Cosmological String Solutions by Dimensional Reduction\thanks{Work
supported by the Department of Energy, contract DE--AC03--76SF00515.}}

\renewcommand{\thefootnote}{\fnsymbol{footnote}}
\vspace{5mm}

\author{
\setcounter{footnote}{2}
K. Behrndt\thanks{e-mail: behrndt@jupiter.slac.stanford.edu, Work
 supported by a grant of the DAAD} \\ {\normalsize \em SLAC, Stanford
University, Stanford, CA 94309}\\
\setcounter{footnote}{6} S. F\"orste
\thanks{e-mail: sforste@vms.huji.ac.il, Work supported by a grant
of MINERVA}\\ {\normalsize \em Racah Institute of Physics, The Hebrew
University, 91904 Jerusalem}}
\date{}

\maketitle

{\em
\begin{center}
Talk presented at \\
``27th Symposium on the Theory of Elementary Particles'' \\
Wendisch--Rietz, September 7--11, 1993
\end{center}
}
\vspace{5mm}

\begin{abstract}
We obtain cosmological four dimensional solutions of
the low energy effective string theory by reducing a five dimensional
black hole, and black hole--de Sitter solution of the Einstein gravity
down to four dimensions.  The appearance of a cosmological constant in
the five dimensional Einstein--Hilbert action produces a special
dilaton potential in the four dimensional effective string action.
Cosmological scenarios implemented by our solutions are discussed.
\end{abstract}

\renewcommand{\arraystretch}{1.6}
\renewcommand{\thefootnote}{\alph{footnote}}
\newcommand{\pa}{\partial}
\newcommand{\be}{\begin{equation}}
\newcommand{\ee}{\end{equation}}
\newcommand{\ba}{\begin{array}}
\newcommand{\ea}{\end{array}}
\newcommand{\aaa}{\alpha}
\newcommand{\eee}{k}
\newcommand{\lll}{\lambda}
\newcommand{\LL}{\Lambda}
\newcommand{\aap}{\alpha^{\prime}}
\newcommand{\sss}{\mbox{$\sigma$}}
\newcommand{\bbb}{\beta}
\newcommand{\vsf}{\vspace{5mm}}

\newpage

In the present paper we describe a procedure how one can obtain new
cosmological solutions by dimensional reduction of a known theory.  We
neglect all terms ${\cal O}(\alpha'^2)$ and consider only curvature
and dilaton terms. Then we can rewrite the effective string action in
four dimensions (4d) as a 5d action without a dilaton. In the construction
of our solutions this 5d theory is our starting point. We take
solutions of this 5d theory, reduce this theory to 4d and interpret
the 4d fields as possible string background fields. Firstly, we want
to discuss only the 5d Einstein-Hilbert action and in a second step we
add a cosmological constant in the 5d theory. Because in 5d we start
with Black Hole (BH) solutions this procedure has the advantage, that
spatially non--flat geometries ($k\not=0$, e.g.\ $S_3$) are
automatically included. The standard approach, i.e.\ to solve the
string equations of motion (vanishing of the $\beta$ functions),
is discussed e.g.\ in \cite{tseyt,venz,perry}. But there the
generalization to $k\not=0$ is rather difficult and a complete
analytical solution for arbitrary $k$ was not yet found.  We should
mention that in the present paper the way through the 5d theory is
used only as a technical tool providing us with a new nontrivial
cosmological solution of pure dilaton gravity. This differs from the
Kaluza--Klein philosophy of \cite{schwarz} where a moduli field
appears as a ``remnant'' of a higher dimensional theory which has its
own physical meaning. 

In the first part of this talk we present the procedure and the solution
\cite{wir} and in a second part we interpret and discuss this solution.

\vsf

So, restricting ourselves on curvature and dilaton terms, the 4d
effective action in the lowest order in the  $\alpha'$ expansion is
given by
\be                                             \label{1}
S=\frac{1}{2}\int d^4 x \sqrt{|G|}\, e^{-2\phi}\left(R +
                    4 (\pa \phi)^2 \right) ~,
\ee
where we assume that the central charge term vanishes. Let us now
transform this 4d action into a 5d Einstein-Hilbert (EH) action.
First we define a five dimensional metric (not depending on the fifth
coordinate),
\renewcommand{\arraystretch}{0.8}
\be                                  \label{2}
G_{\mu\nu}  \rightarrow \tilde{G}_{ab} =
  \left(
     \begin{array}{c|c} e^{\alpha \phi} &\\
                 \hline & \\
                         & \  e^{\beta \phi}\, G_{\mu\nu} \  \\ &
        \end{array} \right)\quad ,\quad  \mbox{with:} \qquad
\alpha = \pm \frac{4}{\sqrt{3}}\quad , \quad
\beta = -2 \left(1 \pm \sqrt{\frac{1}{3}}\right),
\ee
\renewcommand{\arraystretch}{1.6}
and Latin (Greek) indices are running from one to five (four). Using the
five dimensional metric (\ref{2}) and adding one dummy integration
we get for the action (\ref{1}) the  5d EH action
\be \label{3}
S \rightarrow \tilde{S} = \int d^5 x \sqrt{|\tilde{G}|} \, \tilde{R} ~.
\ee
We want to use this procedure for the construction of new cosmological
solutions.
Therefore the 4d metric should be spatially isotropic and homogeneous
(Friedmann-Robertson-Walker),
\be \label{4}
ds^2 = -d\tau^2 + a^2(\tau)\left[ dr^2 + \left(\frac{\sin \sqrt{k}r}{\sqrt{k}}
    \right)^2 \left( \sin^2\theta d\phi^2 + d\theta^2\right) \right]
  = -d\tau^2 + a^2(\tau) d\Omega_{3,k}^{2} .
\ee
The whole dynamics of this metric is contained in the world radius
$a(\tau)$ which has to be determined by the field equations.

First we demonstrate how to get a cosmological solution
for $k=1$. In this case the spatial part of the space time
is a $S_3$ manifold with $a(\tau)$ as the time dependent radius.
The most general 5d metric respecting the $S_3$ symmetry
is given by a Schwarzschild metric which can be written as
\be \label{5}
\tilde{ds}^2 = e^{\nu(t)} dx^2 - e^{\lambda(t)} dt^2 + t^2
d\Omega_{3,k =1}^{2}
\ee
where $x$ is our fifth coordinate which the theory should not depend on
and $t$ corresponds to the time in the 4d theory. In comparison
to the usual Schwarzschild metric our time corresponds to the radius
and $x$ to the time, however, with opposite signs in front of $dx^2$ and
$dt^2$. Since we have no matter in the 5d theory a nontrivial vacuum
solution satisfying the desired $S_{3}$ symmetry is given by a 5d Black Hole.
The generalization of this solution for arbitrary
$k$ is given by
\be            \label{6}
e^{-\lambda} = C \, e^{\nu} = -k + \frac{2m}{t^2} \ .
\ee
After performing the reduction to the 4d theory (2)
we obtain for the dilaton and the 4d metric
\be        \label{7}
\ba{rcl}
\phi&=&\pm\frac{\sqrt{3}}{4} \nu\\
ds^2&=&e^{-\beta \phi} \left( - e^{\lll} dt^2 + t^2 d\Omega_{3,k}^2
\right)\\
  &=&- \left(e^{-\lll}\right)^{\frac{-1\pm\sqrt{3}}{2}} \, dt^2 +
\left(e^{-\lll}\right)^{\frac{1\pm\sqrt{3}}{2}} \, t^2 \,
d\Omega_{3,\eee}^2 ~.
\ea
\ee
Unfortunately, we can perform the integration $(e^{-\lll})^
{\frac{-1\pm\sqrt{3}}{2}} \, dt^2 = d\tau^2$ only numerically and
can not find an analytic expression for (\ref{4}) . At the end we
will discuss some special cases and present some numerical results.
In addition, in order to have a real metric in (\ref{7}) we have
the restriction that (\ref{6}) has to remain positive, i.e.\
$\frac{2m} {t^2} > \eee$ defines the physical $t$ region. Therefore
in general (e.g.\ for $m>0$) the universe starts at $t=0$ and ends
at the zero of (\ref{6}), i.e.\ at the horizon of the 5d theory. If
(\ref{6}) has no zero (for $k=-1,0$) we do not need to
restrict the $t$ region. With horizons we mean always the 5d BH
horizons. They do not correspond to horizons in the 4d theory,
instead, they are singular points namely the end or the beginning of
the universe. The singularity at these points is caused by the Weyl
transformation in (\ref{7}).

Before we discuss the solution (\ref{7}) in detail let us 
describe a possible generalization of the 5d theory. The simplest
extension is given by adding a cosmological constant, i.e.
\be                    \label{8}
\tilde{S} \rightarrow \tilde{S} = \int d^5x \sqrt{|\tilde{G}|}
     \left( \tilde{R} - \LL \right) ~.
\ee
Again we look for a ``static'' BH solution and find for arbitrary
$\eee$
\be                  \label{9}
e^{-\lambda} = C \,e^{\nu} = -\eee + \frac{2m}{t^2} + \frac{\LL}{12} t^2 ~.
\ee
For $\eee = 1$ this solution corresponds to the known 5d
Schwarzschild - de Sitter metric (after interpreting $x$ as time
and $t$ as radius). The constant $C$ can be eliminated by a constant rescaling
of $x$ or equivalently by a constant shift
in the dilaton (cf. (\ref{7})), i.e. the constant part of the
dilaton ($\phi \sim \phi_0 + \phi(t)$) is fixed by the $x$ scale.
Another useful parameterization is given by
\be  \label{10}
e^{-\lll} = \frac{\LL}{12} \frac{(t^2 - t_+) (t^2 - t_-)}{t^2}
  \quad ,\qquad \mbox{with} \qquad t_{\pm} =
 \frac{6 \eee}{\LL} \left(1 \pm \sqrt{1 -
  \frac{2}{3}  \frac{m \LL}{\eee^2}}\right) ~.
\ee
where $t_{\pm}$ are the two horizons (BH and de Sitter) of the
Schwarzschild - de Sitter metric. Both horizons
coincide at the critical limit $3 \eee^2 = 2m \LL_{cr}$.
In order to get a real 4d metric we have again (as for $\LL = 0$)
the restriction that $e^{-\lll} > 0$.
If we reduce the 5d action (\ref{8}) in terms of (\ref{2})
to the 4d theory we see that the cosmological constant produces
a special dilaton potential in four dimensions
\be            \label{11}
\tilde{S} = \frac{1}{2} \int d^5x \sqrt{|\tilde{G}|}
  \left( \tilde{R} - \LL \right) \quad \rightarrow \quad S = \frac{1}{2}
  \int d^4 x \sqrt{|G|}\,e^{-2\phi} \left( R + 4 (\pa \phi)^2 -
  \LL e^{\bbb \phi}  \right) ~,
\ee
where $\beta$ is defined in (\ref{2}).

\vsf

To get contact with standard cosmology let us now consider the
solution (\ref{9}) in the Einstein frame. This frame is defined by
the Weyl transformation: $G_{\mu\nu}^{(E)} = e^{-2\phi} G_{\mu\nu}$
and for the effective action (\ref{11}) we find
\be   \label{12}
\begin{array}{c}
S=\int d^4 x \sqrt{ |G^{(E)}| } \left[ R^{(E)} - 2 (\pa\phi)^2 -
  \LL \, e^{\mp\frac{2}{\sqrt{3}} \phi}    \right] ~.
\end{array}
\ee
The action (\ref{12}) contains the well known Einstein--Hilbert term
describing the gravitational part of the theory. Therefore, the Einstein 
frame is more popular from the point of view of the Einstein gravity.
But there are also some arguments in favour of the string frame
\cite{bellid}, e.g.\ the free motion of a string follows geodesics in
the string frame and not in the Einstein frame.  In the following we
are going to consider both frames on an equal footing. The matter part
of (\ref{12}) is given by the dilaton terms (however, the kinetic term
has the wrong sign) and the corresponding energy--momentum tensor is
\begin{equation}  \label{13}
T_{\mu\nu}^{matter} = 2 \left( \partial_{\mu} \phi \partial_{\nu} \phi
 - \frac{1}{2} (\partial \phi )^2 G_{\mu\nu}^{(E)} \right) -
 \frac{1}{2} \LL \, e^{\mp\frac{2}{\sqrt{3}} \phi} \, G_{\mu\nu}^{(E)} ~.
\end{equation}
In this frame the 4d metric is given by
\be  \label{14}
ds_{E}^2 = - e^{\lll /2} dt^2 + t^2 e^{-\lll /2} d\Omega_{3,\eee}^2
\ee
and we see that the ``$\pm$'' ambiguity of the string metric (\ref{7})
droped out. The origin of this ambiguity is discussed below.

\vsf

\begin{figure}[t] \vspace*{-25mm}
\begin{minipage}[t]{7.3cm}
\mbox{\epsfig{file=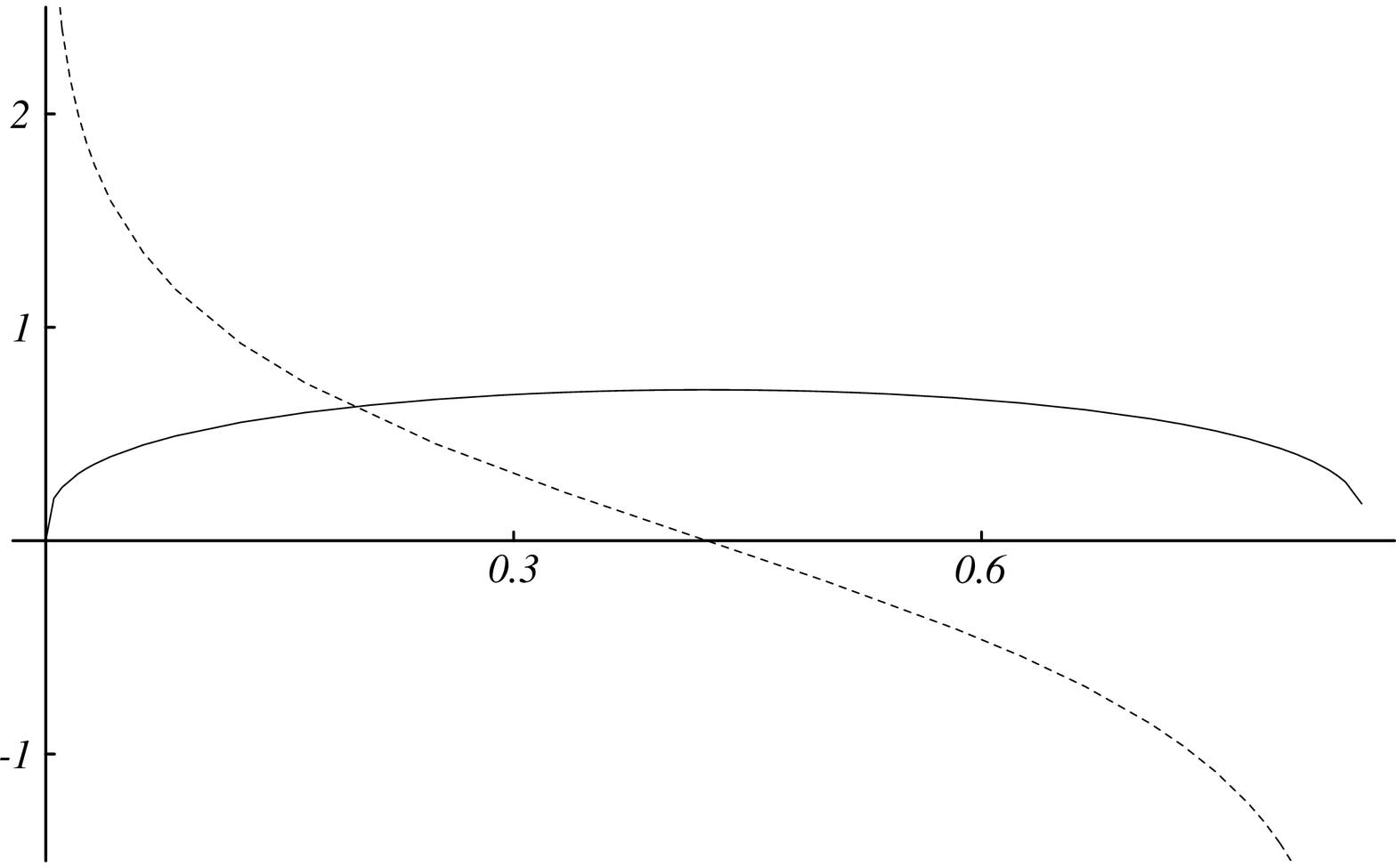,height=10cm,width=7cm}}
      \vspace{-27mm} \par
    \caption{$a(\tau )$ and $\phi(\tau)$ (dashed) in the Einstein frame
   for $k=1$, $m=\frac{1}{2}$, $\LL =0$}
     \end{minipage}
 \hfill
 \begin{minipage}[t]{7.3cm}
 \mbox{\epsfig{file=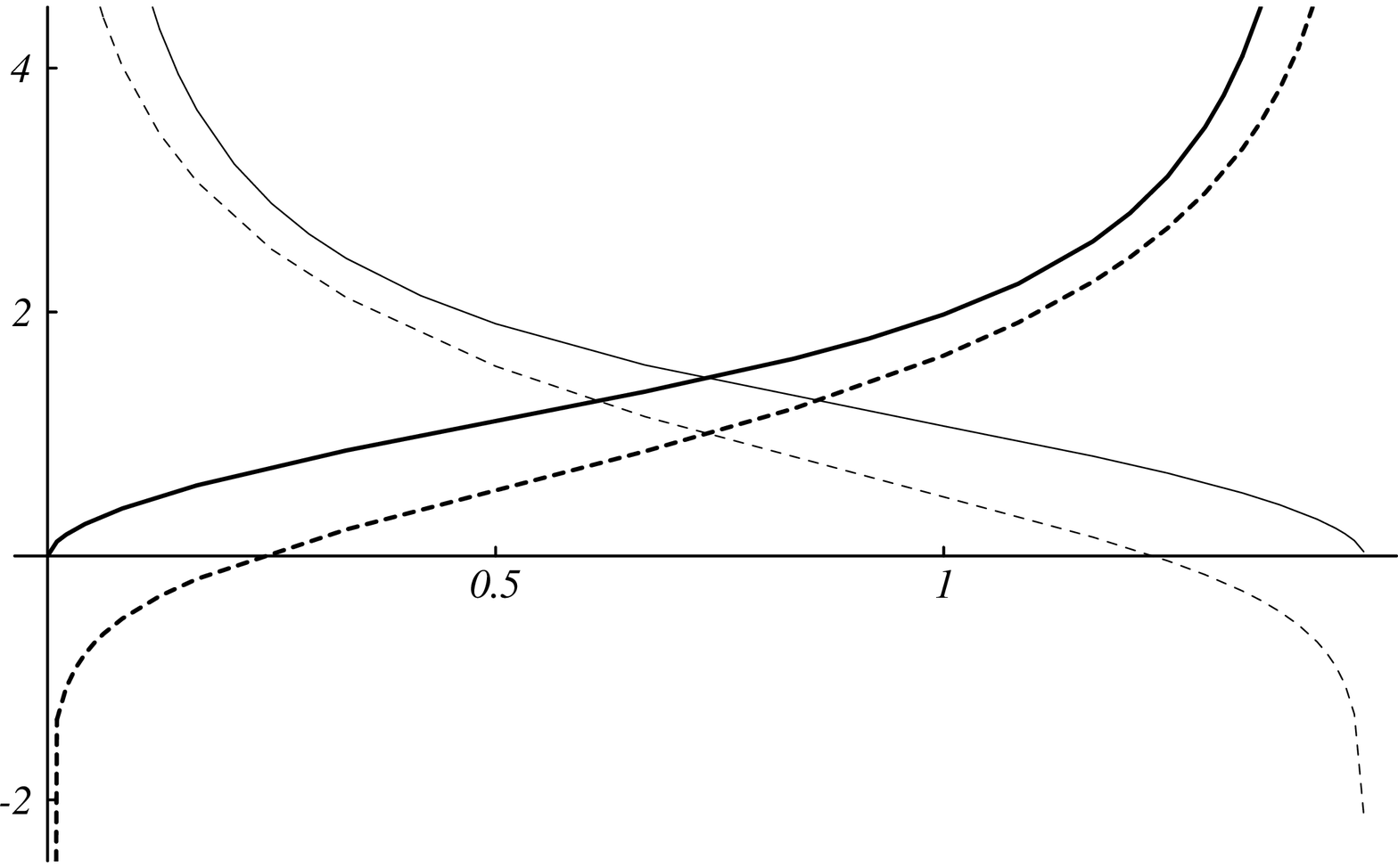,height=10cm,width=7cm}} \vspace{-27mm} \par
\caption{
 $a(\tau )$ and $\phi(\tau)$ (dashed) in the string frame for $k=1$,
 $m=\frac{1}{2}$, $\LL =0$}
 \end{minipage} \vspace{-3cm}
\label{bild1}
\begin{minipage}[t]{7.3cm}
 \mbox{\epsfig{file=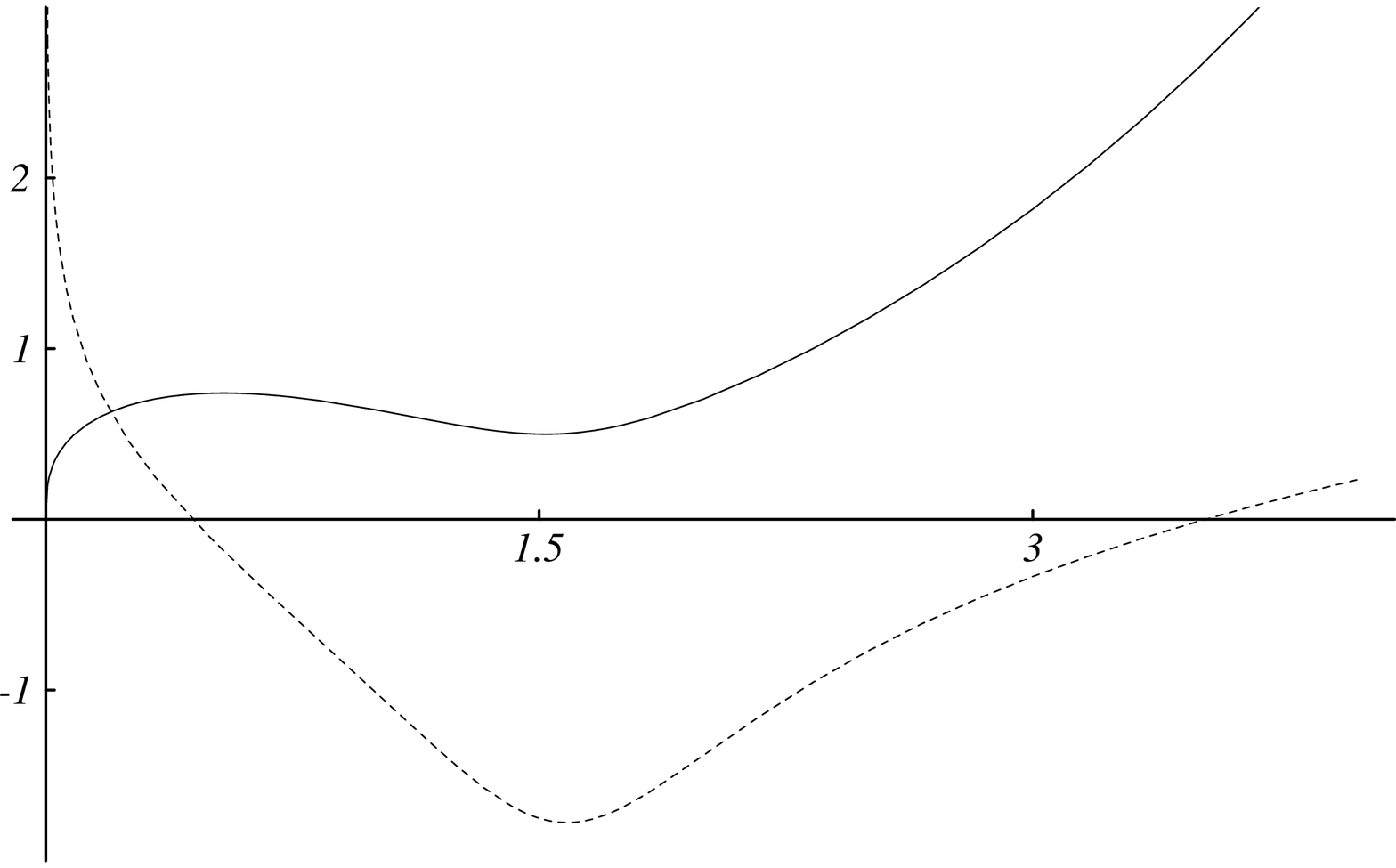,height=10cm,width=7cm}}
 \vspace{-27mm} \par
\caption{
$a(\tau )$ and $\phi(\tau)$ (dashed) in the Einstein frame for $k=1$,
  $m=\frac{1}{2}$, $\LL =3.1$}   \hfill
\end{minipage}
\hfill
\begin{minipage}[t]{7.3cm} \hfill
 \mbox{\epsfig{file=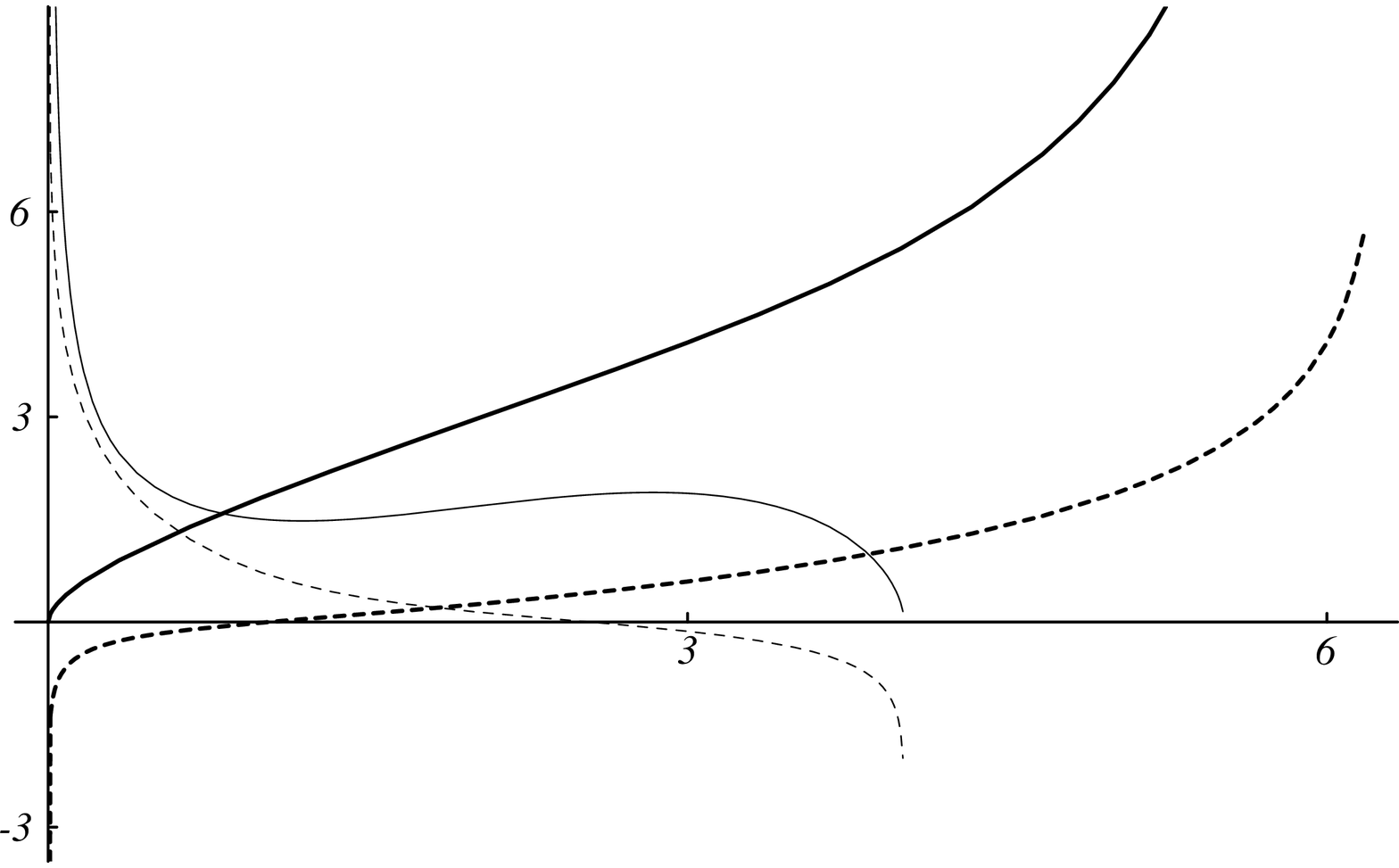,height=10cm,width=7cm}} \vspace{-27mm} \par
    \caption{ $a(\tau)$ and $\phi(\tau)$ (dashed) in the string frame
    for $k=-1$, $m=\frac{1}{2}$, $\LL=-1$  }
\label{bild4}
\end{minipage}\vspace{-10mm}
\end{figure}
In order to obtain statements about the evolution of the (4d) universe
we have to bring the solution (\ref{7}) and (\ref{14})
into the standard form (\ref{4}), where the world radius
$a\left(t(\tau)\right)$ and the function $t(\tau)$ are defined by
\be  \label{15}
\ba{lll}
a^2 = t^2\, (e^{-\lll})^{\frac{1\pm\sqrt{3}}{2}}  \qquad ,  & \qquad
\dot{t}^2(\tau) = (e^{-\lll})^{\frac{1\mp\sqrt{3}}{2}} \qquad  & \qquad
 \mbox{(in the string frame)} ~,\\
a_{(E)}^2 = t^2\, e^{-\lll /2}  \qquad , & \qquad
\dot{t}^2(\tau) = e^{-\lll /2} \qquad & \qquad
 \mbox{(in the Einstein frame)} ~,\\
e^{-\lll} = - \eee +\frac{2m}{t^2} + \frac{\LL}{12} t^2 \quad . & &
\ea
\ee
Unfortunately we were not able to solve these equations analytically.
Instead, we have plotted some numerical results. Figure 1 is an example
for a closed universe
($k=1$) with finite lifetime and vanishing dilaton potential.
The expansion starts at $\tau =0$, reaches the maximum at $t^{2}(\tau ) =m$,
and shrinks to zero size at the ``horizon'' $t^2(\tau )=2m$.
The corresponding ``lifetime'' of the universe is according to (\ref{15})
given by
\be  \label{16}
\tau_0 = \int_0^{\sqrt{2m}} \left(\frac{t^2}{2m -
  t^2}\right)^{\frac{q}{2}} dt =
    \sqrt{\frac{2m}{\pi}}\, \Gamma(\frac{q+1}{2}) \, \Gamma(1-\frac{q}{2})
\ee
where: $q=\frac{1}{2}(1 \mp \sqrt{3})$ in the string frame and
$q = \frac{1}{2}$ in the Einstein frame. The left-right symmetry of
this solution is simple a consequence of the symmetry: $t^2
\leftrightarrow 2m - t^2$ for (\ref{14}).
\mbox{Figure 2} shows the same universe in the string frame. In this
frame we have to take into account the ``$\pm$'' ambiguity, the
thinner line corresponds to the upper sign in (\ref{7}) whereas the
bold line is the solution for the lower sign. In both cases we have
a hyper(de)inflation. This means the universe expands in a finite
proper time $\tau$ from zero to infinity or vice verse (see also
(\cite{perry})).  In Figure 3 we can see the influence of a
non-vanishing dilaton potential. At the beginning it looks similar
to Figure 1, but in the contracting phase the potential
becomes relevant and an accelerated expansion starts. In the last  
figure we have plotted an example for an open universe in the string
frame. For the lower sign the world radius goes from zero to infinity
and for the other sign it is vice verse. Again, the lifetime is
finite.

\vsf

\noindent
Now we want to investigate some special questions in detail.

{\bf 1. Asymptotic behavior:} From (\ref{15}) we find that it is always
possible to fix the integration constant by $t(0)=0$. Then, for
$\tau\rightarrow 0$ (or $t\rightarrow 0$) we get
\begin{equation}       \label{17}
\ba{ccc}
a(\tau ) \sim \tau ^{\mp \frac{1}{\sqrt{3}}} &\quad , \quad &
a_{E}(\tau ) \sim \tau ^{\frac{1}{3}}\\
e^{2\phi}\sim\tau^{-(1\pm \sqrt{3})}     &\quad , \quad &
e^{2\phi_{E}}\sim\tau^{\mp \frac{2}{\sqrt{3}}} ~.
\ea
\end{equation}
Although the dilaton is not transformed if we go to the Einstein frame
we have to perform different time redefinitions in both frames.
For $k =\Lambda =0$ (\ref{17}) is an exact expression for all $\tau$
and coincides with the solution of Mueller \cite{muell}.
The asymptotic behavior at the end (or late times) of the universe
is depending on whether the lifetime is finite or infinite.
 In the second case we find ($\tau \propto
\infty$)
\be                                    \label{18}
\ba{ccc}
a(\tau ) \sim \left\{
\ba{cc}
\tau^{\sqrt{3}} & \LL > 0 \\
\tau & \LL = 0 \ea \right.
 & \quad , \quad &
a_{E} \sim \left\{
\ba{cc}
\tau^{3} & \LL > 0 \\
\tau & \LL = 0 \ea \right. \\
e^{2\phi}\sim \left\{
\ba{cc} \tau^{3 - \sqrt{3}} & \LL > 0\\
1 & \LL =0 \ea \right.
   & \quad , \quad &
e^{2\phi_{E}}\sim \left\{
\ba{cc} \tau^{2\sqrt{3}}& \LL > 0 \\
1 & \LL = 0 \ea \right.  ~.
\ea
\ee
For vanishing potential ($\LL =0$) this limit corresponds to $k =-1$
($e^{-\lambda} > 0$) and we have the remarkable consequence that in
the asymptotic limit our solution is in both frames a flat space time
($K=\tau$) with constant dilaton.  This is a result of the
asymptotic flatness of the 5d BH. In addition, we have here no
``$\pm$'' ambiguity in the string frame, because for the lower sign
the lifetime of the universe is always finite for $\LL\not= 0$ and for
$\LL=0$ one gets for both signs the flat limit. Finally, at the end of
a finite lifetime, i.e.\ near a horizon or for the lower sign in the
string frame for $t\rightarrow\infty$, the asymptotic behavior is
given by
\be   \label{19}
\ba{lcl}
a(\tau ) \sim (\tau_{0} -\tau )^{\pm \frac{1}{\sqrt{3}}} &\quad ,\quad &
a_{E}(\tau ) \sim (\tau_{0} -\tau )^{\frac{1}{3}}\\
e^{2\phi}\sim(\tau_0-\tau )^{-(1\mp\sqrt{3})} &\quad , \quad &
e^{2\phi_{E}}\sim (\tau_0-\tau )^{\pm\frac{2}{\sqrt{3}}}  ,
\ea
\ee
where $\tau \rightarrow \tau_{0}$ ($\tau_0$ is the lifetime of the
universe). If we compare the behavior at the beginning and the end of
the universe we get for finite lifetime the statement, that the world
radius $a(\tau)$ in the string frame always has a singularity, either
at $\tau=0$ or $\tau=\tau_0$, i.e.\ there is a hyper(de)inflation.  On
the other hand $a_E$ remains finite and vanishes at $\tau=0$ and
$\tau=\tau_0$ (see also \cite{venz,perry}). One example for this is
plotted in figure 2.

{\bf 2. Dilaton behavior:} Apart from the flat limit ($\LL=0$, $\tau\propto
\infty$) the dilaton is always 
infinite at the beginning and the end of the universe. The reason is
that these points are given by the zeros of $e^{\lll}$ or $e^{-\lll}$
and the dilaton is: $\phi=\mp\frac{\sqrt{3}}{4} \lll$ (see (\ref{7})).
Thus, at the beginning or the end we have either a strong or weak
string coupling limit ($g_s \sim e^{2\phi}$) and the ``$\pm$''
ambiguity switches between both.  Because $ds_{E}$ (\ref{14}) does not
depend on this ambiguity the Einstein metric is invariant under a change
of the strong to the weak coupling limit and vice verse. Let us now
investigate whether the dilaton has maxima or minima. At these points
the first derivative vanishes and the energy momentum tensor
(\ref{13}) is given by a cosmological constant and we can expect a
phase of exponential expansion.  These points are defined by:
$\dot{\phi}(t(\tau))=
\phi'(t) \, \dot{t}(\tau)=0$. From (\ref{15}) we know that $\dot{t}$
is a non-vanishing function (because: $e^{-\lambda}>0$) and
$\phi'=0$ yields
\be  \label{20}
t^{2}_{extr}=\sqrt{\frac{24 m}{\LL}} 
\ee
or 
\be
e^{-\lll}=-k + \sqrt{\frac{2 \LL m}{3}} \quad , \quad  \phi(t_{extr})=
\pm\frac{\sqrt{3}}{4}\log\left(-k + \sqrt{\frac{2 \LL m}{3}}\right)
\ .
\ee
Since in the physical region $e^{-\lll}$ has to be positive
and finite we have the result that the dilaton has maxima or minima
only if: $\LL > \LL_{cr}=\frac{3}{2 m}$ for $k=1$ or $\LL > 0$ for
$k=-1,0$.  All these cases are only fulfilled if in the 5d theory are
no horizons and therefore the $t$-region is infinite. Furthermore,
since the second derivative at this point does not vanish, this
extremum is the global maximum or minimum of the dilaton.  Thus we
have the result: the dilaton has at most one extremum and in this case
the dilaton is $+ \infty$ (strong coupling limit) at the beginning {\bf and}
at the end of the universe or the dilaton is $- \infty$ (weak coupling
limit) at both ends of the evolution of the universe.
Figure 3 shows an example.

{\bf 3. ``$\pm$'' ambiguity:} Mathematically this ambiguity appeared at
the reduction of the 5d Einstein--Hilbert theory to the 4d string
effective action. But physically there is another possible interpretation.
If we set the cosmological constant to zero the 5d theory can
also be considered as a string theory with vanishing dilaton. $\LL=0$ is
necessary in order to get a vanishing dilaton $\beta$ function.
In contrast to our 4d theory the 5d theory posses an abelian isometry
corresponding to the independence of the fifth coordinate. This has the
consequence that we can construct a new 5d string solution by a duality
transformation \cite{busch}. Our 5d Einstein--Hilbert action then becomes
\be \label{21}
\int d^5 x \sqrt{\tilde{G}} \tilde{R} \rightarrow \int d^5 x
\sqrt{\bar{G}} e^{-2\psi}\left(\bar{R} + 4 (\partial\psi)^2\right)
\ee
and the new 5d solution is connected with the old one by
\be \label{22}
\bar{G}_{55}=\frac{1}{\tilde{G}_{55}}  \qquad , \qquad
\psi = -\frac{1}{2}\log \tilde{G}_{55} \ ,
\ee
where $\psi$ is now a 5d dilaton field. If we now reduce this dualized
5d string theory to a 4d string theory we get again the solution
(\ref{7}) but with switched ``$\pm$'' signs. Thus, for $\LL = 0$
both solutions (\ref{7}) are connected by dualizing of the 5d theory
and both solutions can be generalized to $\LL\not= 0$. In addition,
for $k=0$ it is possible to relate both solutions by dualizing
the spatial isometries \cite{ven2}. Physically this duality transformation
changes the sign of the dilaton and therewith switches between the
strong and weak coupling limit at the beginning or the end of the
universe. In the string frame both solutions are physically different
(see figure 2)
whereas the 4d Einstein metric is invariant under this transformation
(e.g.\ figure 1).

\vsf

To summarize, in the present paper we have obtained various
cosmological solutions (\ref{7}) of the low energy effective string action.
Although the method presented is very simple (reduction of a
5d Black Hole solution) our solution has as far as we know not been
obtained before.  The reason is that  we did not use the standard
parameterization of the Robertson-Walker metric (\ref{4}) and it
seems to be impossible to solve the string equations of motion
(vanishing of the $\beta$ function) for this coordinate system
analytically. In our procedure this becomes manifest in the failure
to find an analytic solution of (\ref{15}). However, in
most cases it is possible to get some impression about the features of
our solutions. We have investigated the asymptotic behavior of the
world radius near zero, near the horizons of the corresponding 5d
Black Hole, and in the infinite future. For vanishing dilaton
potential and $k = -1$ we obtained a flat universe in the large time
limit.  In the case where we were able to get an analytic expression
($\LL = k =0$) in the Robertson-Walker parameterization (\ref{4})
our result coincides with Mueller's solution \cite{muell}. Furthermore,
the dilaton is divergent at the beginning and at the end of the universe
and a duality transformation in the 5d theory switches between the
strong ($\phi \rightarrow +\infty$) and the weak ($\phi \rightarrow
-\infty$) string coupling limit. While the 4d string frame metric
is not invariant under 5d duality the 4d Einstein frame metric
is invariant.

The procedure developed in this paper should be applicable to more
general 5d theories. In this paper we have discussed the most
easiest case of pure dilaton gravity. The inclusion of further
background fields like antisymmetric tensor and gauge field
in the 5d theory is in progress.

\newpage

\noindent
{\large\bf Acknowledgments} \vspace{3mm} \newline \noindent
We would like to thank H.\ Dorn, D.\ L\"ust and G.\ Weigt
for helpful discussions.

\vspace{8mm}

\noindent
{\em Note added in proof.} As this paper was completed, we received 
some references \cite{wltsh} in which cosmological solutions 
as interior part of black hole solutions have been considered, too.
For drawing our attention to these papers we are grateful to
D.L. Wiltshire.

\renewcommand{\arraystretch}{1.0}

\end{document}